\documentclass[aps,prb,twocolumn,showpacs,preprintnumbers,amsmath,amssymb,superscriptaddress]{revtex4-1}

\usepackage{graphicx}
\usepackage{amssymb}
\usepackage{epstopdf}
\usepackage{dcolumn}
\usepackage{bm}
\usepackage{amsmath}
\usepackage{mathrsfs}
\usepackage{siunitx}
\usepackage{subfigure}
\usepackage{flafter}
\usepackage{caption}
\usepackage{tabularx}
\usepackage{float}
\usepackage{braket,mleftright}
\usepackage[font={small}]{caption}
\newcolumntype{C}[1]{>{\centering\let\newline\\\arraybackslash\hspace{0pt}}m{#1}}

\begin{document}

\title{Proximity induced quantum anomalous Hall effect in graphene/EuO hetero-structures} 

\author{Shanshan Su}
\email{ssu008@ucr.edu}
\affiliation{Department of Electrical and Computer Engineering, University of California, Riverside, CA 92521, USA}
\affiliation{Center of Spins and Heat in Nanoscale Electronic systems, University of California, Riverside, CA 92521, USA}

\author{Yafis Barlas}
\email{yafisb@gmail.com}
\affiliation{Department of Physics and Astronomy, University of California, Riverside, CA 92521, USA}
\affiliation{Center of Spins and Heat in Nanoscale Electronic systems, University of California, Riverside, CA 92521, USA}

\author{Roger K. Lake}
\email{rlake@ece.ucr.edu}
\affiliation{Department of Electrical and Computer Engineering, University of California, Riverside, CA 92521, USA}
\affiliation{Center of Spins and Heat in Nanoscale Electronic systems, University of California, Riverside, CA 92521, USA}
\date{\today}

\begin{abstract}
In a heterostructure of graphene and the ferromagnetic insulator EuO, 
the Eu atoms induce proximity exchange and
inter-valley interactions in the graphene layer. 
Constrained by the lattice symmetries, and guided by {\it ab initio} calculations, 
a model Hamiltonian is constructed that describes the low-energy 
bands.
Band parameters such as proximity induced exchange splitting, 
spin orbit coupling, and inter-valley interaction are determined. 
Calculations of the Chern number identify the 
conditions under which the hetero-structures exhibit topologically non-trivial bands
that give rise to the quantum anomalous Hall effect
with a Hall conductivity of $\sigma_{xy} = 2 e^2/h$. 
\end{abstract}

\maketitle 


\section{Introduction}

Ever since the classification of the integer quantum Hall effect (IQHE) 
in terms of topological invariants~\cite{TKNN}, 
significant theoretical effort has gone towards realizing IQHE phenomenology at vanishing external 
magnetic fields. 
Haldane\citep{HaldaneQAH} proposed that in the presence of an intrinsic spin-orbit coupling, 
spinless electrons hopping on a two-dimensional honeycomb lattice are topologically 
non-trivial which can result in one-dimensional chiral modes along the edges of an insulator. 
Owing to the chiral nature of these edge modes, 
the gapless edge states are dissipationless and exhibit a Hall conductance $\sigma_{xy} = e^2/h$. 
This Hall conductance is a consequence of the Berry curvature 
associated with the Bloch bands in momentum space, 
and it is quantized only when the Fermi energy lies in the bulk band gap of the material. 
A number of other proposal have been made for realizing
the quantum anomalous Hall (QAH) effect in 
mercury-based quantum wells\cite{HgQAH}, 
optical lattices\cite{CWQAH}, 
disorder induced Anderson insulators\cite{disorderedAIQAH}, 
magnetic topological insulators \cite{yu_quantized_2010,jiang_quantum_2012}, 
and ferromagnetic graphene\cite{qiao_quantum_2010,Niu_QAH_longpaper}. 
The robust nature of charge transport which identifies the QAH effect at vanishing magnetic fields might enable 
design of novel quantum devices for low-power electronics applications. 
\begin{figure*}[th]
\centering
\includegraphics[width=7.0in]{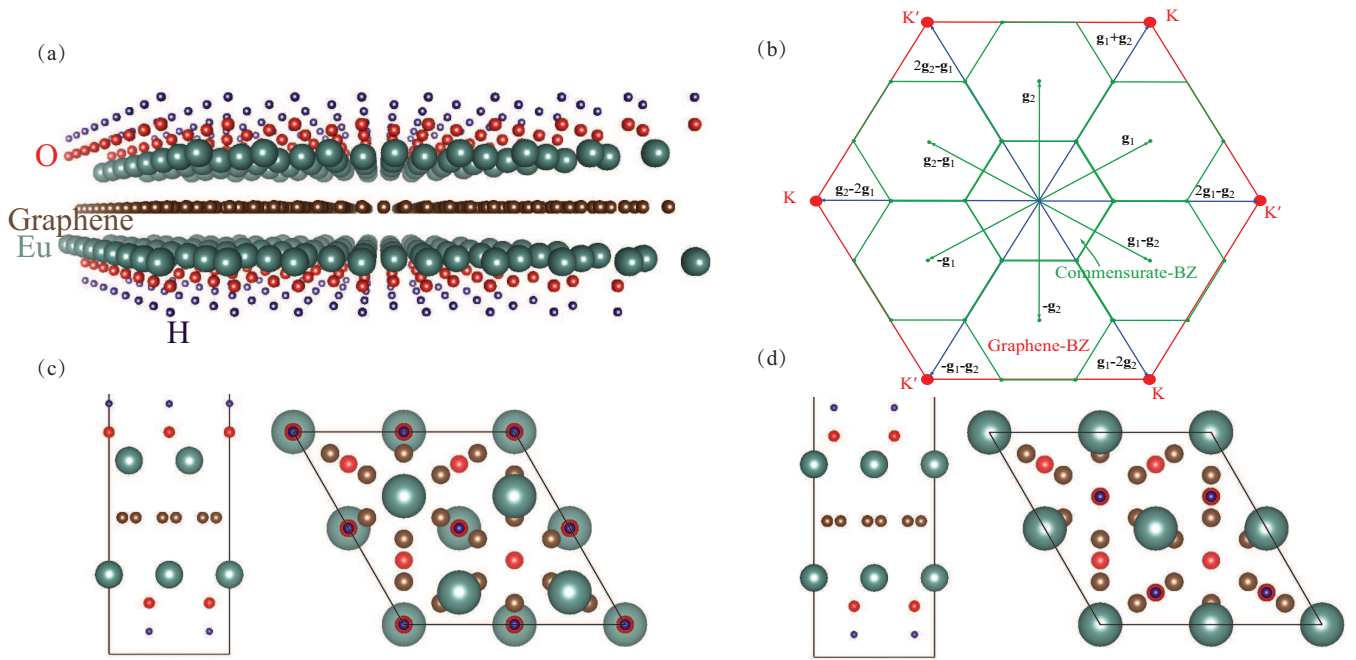}
\caption{(Color online) 
(a) Schematic view of a hetero-structure with graphene between two EuO layers. 
The O layers are terminated with H atoms, and the surfaces abutting the graphene are the 
Eu (111) planes. 
(b) The reciprocal lattice corresponding to the unit cells shown in (c) and (d)
maps the ${\bf K}$ and ${\bf K}'$ points of the hexagonal graphene Brillouin Zone (BZ) 
indicated by the outer red hexagon to the ${\bf \Gamma}$ point of the 
commensurate BZ of the graphene/EuO unit cell indicated by the central green hexagon.
The ${\bf{g}_{i}}$'s are the reciprocal lattice vectors of the hetero-structure unit cells in (c) and (d).  
Elevation and plan views of the unit cells corresponding to the two graphene/EuO 
geometries are shown in (c) for the Eu-misaligned structure  
and (d) for the Eu-aligned structure.
}
\label{fig:combined1} 
\end{figure*}

To realize the QAH state in realistic materials two conditions are necessary, 
(i) broken time reversal symmetry and 
(ii) topologically non-trivial bands. 
Since topological insulators (TIs) possess a large spin-orbit 
coupling~\cite{TIreview}, a route towards realizing the QAH effect is to introduce ferromagnetic ordering in TIs. 
Immediately following recent successes in synthesizing magnetic TIs (MTIs), transport measurements in 
MTIs verified the predicted $e^2/h$ Hall conductance~\cite{yu_quantized_2010,kou_scale-invariant_2014}. 
Another approach is to engineer the QAH state in ferromagnetic graphene 
in the presence of Rashba spin-orbit coupling \cite{qiao_quantum_2010,Niu_QAH_longpaper}. 
Recently, graphene was successfully deposited on an atomically thin-film insulating ferrimagnet (YIG),
and the transport measurements revealed an  unquantized anomalous Hall effect due to
proximity induced ferromagnetism~\cite{ShiPRL}.
%
%
Several other magnetic material/van der Waals (vdW) materials combinations 
(for example graphene/EuO, graphene/BiFeO$_3$ and MoTe$_2$/EuO \cite{qiao_quantum_2010, qiao_quantum_2014, 
yang_proximity_2013, qi_giant_2015}) 
have been proposed for possible spintronics \cite{qiao_quantum_2014,yang_proximity_2013} 
and valleytronics \cite{qi_giant_2015} applications. 
In these systems, ferromagnetic ordering is induced by a proximity effect, 
which is an efficient way to make a non-magnetic two-dimensional crystal magnetic. 
Additionally, proximity induced magnetism by a magnetic insulator
allows for control of the electron and hole densities by gating. 
Previous theoretical studies have reported an exchange splitting gap of 36 meV in 
graphene/EuO hetero-structures~\cite{yang_proximity_2013}. 

In this paper, we investigate the possibility of a QAH effect in a EuO/graphene/EuO hetero-structure 
as shown in  Fig.  \ref{fig:combined1}.
%
%
The hetero-structure symmetries guide our
construction of an effective Hamiltonian that captures the low-energy band 
dispersion.  
{\it Ab initio} calculations provide parameters such as exchange 
splitting and spin-orbit coupling. 
The model Hamiltonian captures the influence of the Eu atoms 
which leads to two distinct types of inter-valley interactions in addition to the exchange interactions 
and Rashba spin-orbit coupling. 
An analysis of the graphene/EuO model Hamiltonian 
provides conditions under which the systems acquire a non-zero Chern number
and exhibit a Hall conductance of $2e^2/h$. 

The rest of the paper is organized as follows. 
Section II presents the graphene/EuO hetero-structures, the 
density functional theory (DFT) calculations, 
and the band dispersions resulting from the {\it ab initio} calculations. 
Section III describes
the model Hamiltonian that captures the low-energy band dispersion 
of the graphene/EuO hetero-structures with and without spin-orbit coupling.
In section IV,  
calculations of the Chern numbers identify the conditions 
which yield topologically non-trivial bands and
the QAH effect.
Section V concludes with a discussion of the possibility of proximity induced exchange 
and the observation of QAH effect in other graphene/ferromagnetic hetero-structures. 

\section{First principle calculations}
The band dispersions of the EuO/graphene/EuO hetero-structures 
are calculated using the Vienna {\it ab initio} simulation package (VASP)\citep
{kresse_efficient_1996,kresse_ab_1993,kresse_efficiency_1996} in the projected-augmented-wave method \citep
{blochl_projector_1994}. 
The generalized gradient 
approximation (GGA) of the Perdew-Burke-Ernzerhof form \citep{perdew_atoms_1992, wang_correlation_1991, 
kresse_ultrasoft_1999} 
is used for the exchange correlation energy, 
and a Hubbard-U correction is used for the magnetic insulator, EuO. 
The on-site Coulomb repulsion and exchange interactions on the 
Eu atom ${4f}$ orbital are 8.3 eV and 0.77 eV, respectively, 
and on the O atom ${2p}$ orbital, they are
4.6 eV and 1.2 eV, respectively \citep{ingle_influence_2008}.
The kinetic energy cutoff is 520 eV for all calculations.
During all structural relaxations, the convergence tolerance on the 
Hellmann-Feynman forces is less than 0.03 eV \AA. 
An $8\times 8\times 8$ Monkhorst-Pack k-point mesh 
is used for bulk EuO. 
The calculated bulk lattice constant 
is 5.186 {\SI{}{\angstrom}} which is very close to the previously published first principle 
calculations\citep{yang_proximity_2013} and consistent with the experimental results.
The lattice constant $a_0$ of graphene is 2.46 \AA. 
This results in a lattice mismatch of less than 1\% in the $3a_0 \times 3a_0$
unit cells shown in Figs. \ref{fig:combined1}(c,d).

\begin{figure}
\centering
\includegraphics[width=.5\textwidth]{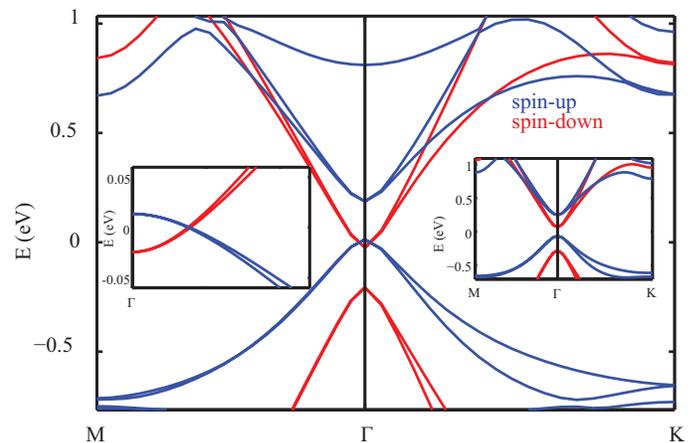}  
\caption{ 
(Color online)
Band structure of the Eu-misaligned structure without SOC. 
Left inset: close-up of the low-energy band structure of the misaligned structure 
near ${\bf \Gamma}$. 
Right inset: Band structure of the Eu-aligned structure.}
\label{fig:combined2} 
\end{figure}

The EuO/graphene/EuO structures consist of 
graphene between the (111) Eu planes of EuO. 
The hetero-structure with graphene on the Eu-terminated surface is 
more stable than graphene on the O-terminated surface \citep{yang_proximity_2013}. 
The relaxation of the 2D hetero-structures uses the same level of theory, cutoffs, and tolerances
as described in the previous paragraph with a  Monkhorst-Pack k-point grid of $4\times 4\times 1$. 
A vacuum buffer space over 25 {\SI{}{\angstrom}} is included to prevent 
interaction between adjacent slabs and hydrogen atoms passivate 
the outer oxygen layers of the EuO films.
The relaxed vertical spacing between the Eu and C layers is 2.517 {\SI{}{\angstrom}}
for the misaligned structure of Fig. \ref{fig:combined1}(c)
and 2.555 {\SI{}{\angstrom}}
for the aligned structure of Fig. \ref{fig:combined1}(d). 
These distances are close the value of 2.57 {\SI{}{\angstrom}}
found previously for a single-sided
heterostructure of graphene on EuO \cite{yang_proximity_2013}.

Figs. \ref{fig:combined1}(c,d) show the two different 
hetero-structures considered.
They differ by the alignment of the EuO-monolayer on opposite sides of the graphene 
layer. 
In both cases, graphene is placed on the (111) surface of EuO. 
This gives a commensurate hetero-structure with a lattice constant 
2 times the lattice constant of a EuO unit cell and 3 times that of the 
graphene unit cell.
%
%
In the aligned structure, 
shown in Fig. \ref{fig:combined1}(d), 
the top EuO-monolayer is directly above the bottom EuO-monolayer, 
whereas in the misaligned structure shown in Fig \ref{fig:combined1}(c), 
the top Eu-monolayer has an in-plane displacement of 1.22 
{\SI{}{\angstrom}} with respect to the bottom EuO layer. 
In both structures the Eu atoms either sit at the center of the hexagonal 
graphene unit cell or at the the bridges of the C-C bonds coinciding with the inversion 
symmetric points of graphene's honeycomb lattice. 
Therefore, in-plane inversion symmetry is preserved for both cases. 
However, as a result of the lateral displacement of the EuO layer in the 
misaligned hetero-structure, inversion symmetry perpendicular to the graphene sheet is broken 
in contrast to the aligned hetero-structure where this symmetry is preserved. 
These symmetries play an 
important role in determining the band dispersion and the 
model Hamiltonian of the graphene/EuO hetero-structure.

\subsection{Band dispersion without SOC}

Fig. \ref{fig:combined2} shows the calculated band dispersion 
in the absence of SOC for the 
misaligned structure of Fig. \ref{fig:combined1}(c), 
and the right inset shows the band dispersion
for the aligned structure of Fig \ref{fig:combined1}(d). 
Both band dispersions are calculated along the path 
${\bf M} - {\bf \Gamma} - {\bf K}$ 
of the commensurate BZ of the $3N \times 3N$ graphene lattice.
${\bf K} = (2\pi/3a,2\sqrt{3}\pi/3a,0)$, 
${\bf M} = (0,\frac{2\pi}{\sqrt{3}a},0)$, and $a = 7.38$ {\SI{}{\angstrom}} is the 
lattice constant of the hetero-structure unit cell. 
The most striking difference in the two band dispersions is the presence of a gap 
between the spin-resolved bands of the aligned hetero-structure, 
whereas in the misaligned hetero-structure the spin-up and spin-down bands intersect. 
The calculated values of gaps for both hetero-structures are tabulated in Table \ref{tab:EuO_G_EuO}. 
The energy band gap is denoted by $E_G$. 
The gap between the spin-up electron band and spin-up hole 
band is $\Delta_{\uparrow}$, and the gap between spin-down bands is $\Delta_{\downarrow}$. 
The spin-splitting of the electron and hole bands are $\delta_e$ and $\delta_h$, respectively. 
In Table \ref{tab:EuO_G_EuO} and Fig. \ref{fig:combined2}, 
the positive value of $ E_G = 127$ meV indicates a band gap between conduction and valence band, 
whereas the negative value of $E_{G} = -38$ meV indicates 
a spin resolved band overlap. 
The values of $\Delta_{\uparrow}$ and $\Delta_{\downarrow}$ in the misaligned structure 
are half of their values in the aligned structure. 
Another striking feature of the calculated band dispersion is that the 
low-energy bands of the combined hetero-structures appear at ${\bf \Gamma}$ and have curvature. 
In contrast, to the {\it ab initio} studies of a 
graphene/BiFeO$_3$ hetero-structure \citep{qiao_quantum_2014}, 
the Dirac cones are no longer at the {\bf K} and {\bf K}' points, but at $ {\bf \Gamma} $, 
consistent with earlier first principle studies of graphene/EuO hetero-structures \cite{yang_proximity_2013}. 
This is due to band folding.

\begin{table*}
\begin{centering}
\begin{tabular}{|c|c|c|c|c|c|}
\hline 
Structure & $E_G$ ($meV$) & $\Delta_{\uparrow}$ ($meV$) & $\Delta_{\downarrow}$ ($meV$) & $\delta_e$ ($meV$) & $\delta_h$ ($meV$) \\
 \hline
	Eu aligned & 127 & 309 & 344 & 182 & 217 \\ \hline 
	Eu misaligned & -38 & 173 & 182 &  211 & 220 \\ \hline
\end{tabular}
\captionsetup{width=2\columnwidth}
\caption{\label{tab:EuO_G_EuO} Energy gaps of the EuO-graphene-EuO structures at the Dirac point. 
$E_G$ is the bandgap of the gapped Dirac cone. 
$\Delta_{\uparrow}$ is the spin-up gap, and $\Delta_{\downarrow}$ is the spin-down gap.
The spin-splitting of the electron and hole bands at ${\bf \Gamma}$ are $\delta_e$
and $\delta_h$, respectively.
}
\end{centering}
\end{table*} 
  
The lattice constants of the graphene-EuO unit cell are exactly 
three times those of the graphene unit cell. 
Hence, the reciprocal lattice constant of the commensurate BZ is $\frac{1}{3}$ that of 
graphene's BZ as shown Fig. \ref{fig:combined1}(b). 
The outer hexagon (red - online) is the BZ of the graphene primitive cell, 
and the central hexagon (green - online) is the BZ of the hetero-structure unit cell. 
Fig. \ref{fig:combined1}(b) shows that the ${\bf K}$ and ${\bf K}$' 
points of the graphene BZ 
lie at equivalent ${\bf \Gamma}$ points in the extended zone of the hetero-structure BZ.
This results in zone folding of graphene's 
${\bf K}$ and ${\bf K}'$ points to ${\bf \Gamma}$.
The differences in the dispersions and non-linearity are due to inter-valley coupling. 
We address this feature when we 
construct the model Hamiltonian to describe the band dispersions of the two hetero-structures in section III.  
\begin{figure}
\includegraphics[width=.5\textwidth]{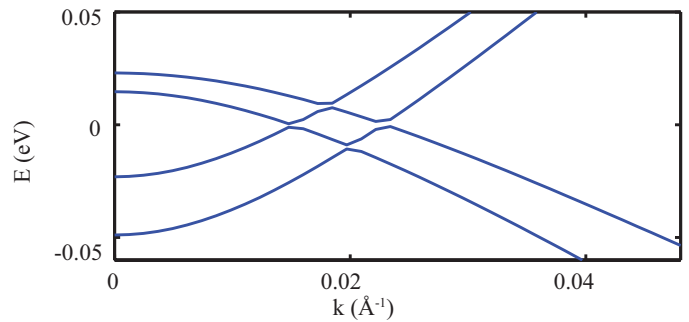}  
\caption{ 
(Color online)
Band structure with SOC of Eu-misaligned structure calculated along the path ${\bf \Gamma}$ to ${\bf K}$ where ${\bf K}$ is 0.57 {\SI{}{\angstrom}}$^{-1}$ away from ${\bf \Gamma}$.}
\label{fig:SOC} 
\end{figure}

\subsection{Band dispersion with SOC}

Our {\it ab initio} calculations that include SOC show very different behaviors 
of the spin resolved bands in the two hetero-structures. 
Since the spin-resolved bands intersect in the misaligned hetero-structure, 
the addition of spin-orbit coupling in this system will be more pronounced than in 
the aligned hetero-structure. 
In the aligned hetero-structure spin-orbit coupling leads to a small splitting of 
the spin resolved bands and the dispersion remains 
gapped, therefore, we focus on the effect of spin-orbit coupling in the misaligned hetero-structure. 

In Fig. \ref{fig:SOC}, we plot the band dispersion of the 
Eu-misaligned hetero-structure along path ${\bf \Gamma} - {\bf K}$. 
Fig. \ref{fig:SOC} shows that SOC breaks the degeneracy of the bands shown
in the left inset of  Fig. \ref{fig:combined2} and gaps the bands.
The conduction bands and the valence bands split by 8 meV and 26 meV at the ${\bf \Gamma}$ point, 
respectively. 
There are two local minimum gaps between the conduction band and valence band
in the band dispersion near ${\bf \Gamma}$ 
with values of 0.1 and 0.3 meV, respectively.
The gap between 
the two conduction bands is 0.2 meV; while the gap between the two valance bands is 1.2 meV. 
The resultant bands show interesting topology on either side of the ${\bf \Gamma}$ point which we address using a 
model Hamiltonian in the next section.

\begin{figure*}[t]
\includegraphics[width=1\textwidth]{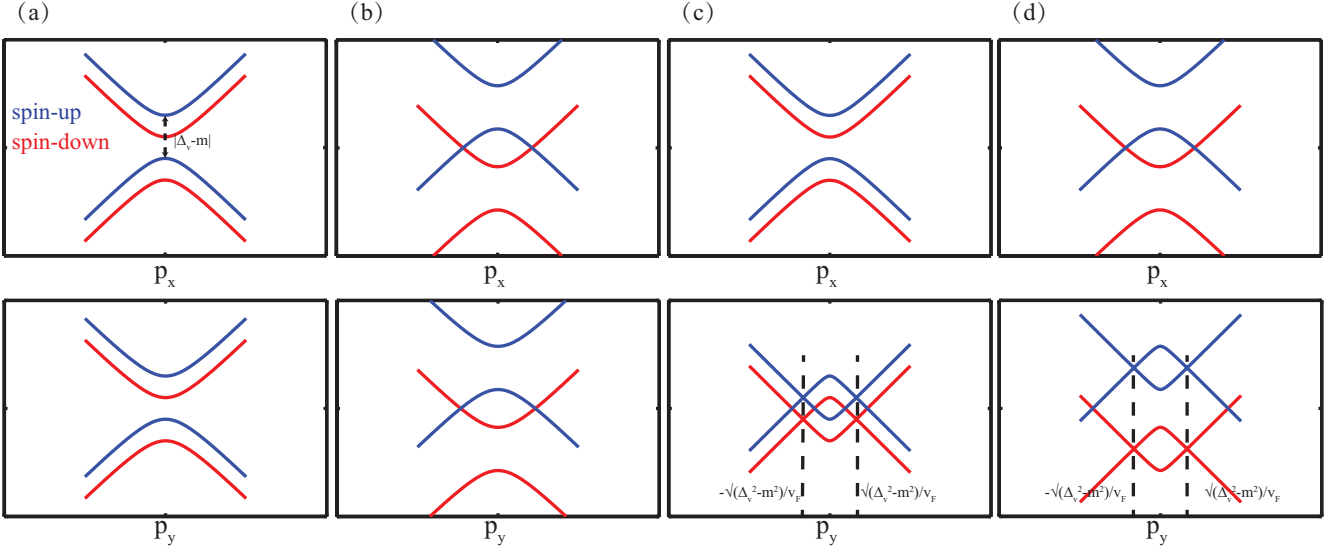}  
\caption{ 
(Color online)
Band dispersions in the absence of SOC for different values of m, $\Delta_{ex}$, and $\Delta_v$. 
(a) $m>\Delta_v$ and $\Delta_{ex}<|\Delta_v-m|$, 
(b) $m>\Delta_v$ and $\Delta_{ex}>|\Delta_v-m|$, 
(c) $m<\Delta_v$ and $\Delta_{ex}<|\Delta_v-m|$, and 
(d) $m<\Delta_v$ and $\Delta_{ex}>|\Delta_v-m|$.}
\label{fig:Configuration_MORE} 
\end{figure*}
 
\begin{table*}
\begin{centering}
\begin{tabular}{|c|c|C{3cm}|C{3cm}|}
\hline 
Relation between $m$ and $\Delta_v$ & value for $\Delta_{ex}$ & band dispersion along $p_x$ & band dispersion along $p_y$ \\
\hline
\hline
$m>\Delta_v$ & $\Delta_{ex}<m-\Delta_v$ in Fig. \ref{fig:Configuration_MORE}(a) &  parabolic, spin splitting & parabolic, spin splitting \\ \cline{2-4} 
 & $m-\Delta_v<\Delta_{ex}$ in Fig. \ref{fig:Configuration_MORE}(b) & parabolic with an overlap between spin-up and spin-down bands & parabolic with an overlap between spin-up and spin-down bands\\ 
\cline{2-4} 
 \hline
 \hline
 $m<\Delta_v$ & $\Delta_{ex}<|m-\Delta_v|$ in Fig. \ref{fig:Configuration_MORE}(c) & parabolic with spin splitting & two Dirac cones with spin splitting \\ \cline{2-4} 
 & $\Delta_{ex}>|m-\Delta_v|$ in Fig. \ref{fig:Configuration_MORE}(d) & parabolic with an overlap between spin-up and spin-down bands & a larger overlap between the spin-up and spin-down double Dirac cones \\ 
\cline{2-4} 
 \hline

\end{tabular}
\captionsetup{width=2\columnwidth}
\caption{\label{tab:Parameter_relation} Description of band dispersions 
in the absence of SOC
for different parameters of Eq. (\ref{eq:NonSOC_H}). 
Plots of the dispersions corresponding to different relative strengths of the model 
parameters are shown in Fig. \ref{fig:Configuration_MORE} 
}
\end{centering}
\end{table*} 

\section{Low-energy effective Hamiltonian}

In graphene, the gapless Dirac cones at {\bf K} and {\bf K'} 
are protected by time-reversal and inversion symmetry. 
Since these Dirac points are separated in the BZ, small perturbations cannot lift this valley 
degeneracy. 
Therefore, the valley index is a good quantum number. 
In the $3 N \times 3 N $ unit cell, due to zone folding of graphene's BZ, 
both valleys {\bf K} and {\bf K}' get mapped to ${\bf \Gamma}$. 
Hence, valley symmetry is no longer 
preserved and inter-valley interactions can gap the Dirac bands at ${\bf \Gamma}$
without breaking inversion or time-reversal symmetry. 
In the graphene/EuO hetero-structures, Eu adatoms positioned at the 
bridge and hollow sites contribute two distinct inter-valley interaction terms
that are responsible for the non-linear dispersions obtained from the {\it ab initio} calculations. 
In this section, we construct a model Hamiltonian that captures the 
effect of these inter-valley interaction terms, 
and we analyze their effect on the band dispersion.

\subsection{Inter-valley interactions}

The following model Hamiltonian that acts on an 8 component spinor is consistent
with the lattice symmetries, and it
describes the salient features of the band dispersion near the ${\bf \Gamma}$ point
in the absence of SOC. 
\begin{equation} \label{eq:NonSOC_H}
H_0 = v_{F} (\hat{\sigma}_x \hat{\tau}_z p_x+ \hat{\sigma}_y p_y)+ \Delta_{ex} \hat{s}_z+ \Delta_v \hat{\tau}_x + m \hat{\sigma}_x \hat{\tau}_x
\end{equation}
In Eq. (\ref{eq:NonSOC_H}), $\hat{\tau}_i$, $\hat{\sigma}_i$ and $\hat{s}_i$ 
are the standard Pauli matrices acting on the valley, 
sublattice, and spin degree of freedom, respectively. 
The first term is the standard low-energy Hamiltonian describing the linear dispersion of the 
Dirac bands in graphene at the two valleys $\tau_{z} = \pm 1$ 
that are now folded to ${\bf \Gamma}$. 
The second term is the exchange coupling term induced by 
the magnetic moment of the Eu atom resulting in proximity induced 
exchange splitting $\Delta_{ex}$ between the spins. 
The last two terms of 
Eq. (\ref{eq:NonSOC_H}) capture the influence of the Eu atoms on the graphene layer. 
In both the hetero-structures of Fig. \ref{fig:combined1}(c,d),
Eu atoms can sit on a C-C bond, referred as the bridge site, 
and in the middle of the hexagon, referred as the hollow site. 
The position of the bridge Eu atom reduces the graphene lattice symmetry from 
$C_{3v} \to C_{2v}$ resulting in the term $\Delta_v \tau_x $ in Eq. (\ref{eq:NonSOC_H}). 
This term corresponds to a valley pseudospin Zeeman term in 
x-direction \citep{ren_single-valley_2015} and shifts the Dirac cones 
from ${\bf \Gamma} =(0,0)$ to $(0,\pm \Delta_{v}/m)$. 
The last term, 
$m \hat{\sigma}_x \hat{\tau}_x$ results from the Eu atom sitting at the hollow site of graphene hexagon; 
we refer to it as an inter-valley scattering term. 
The combined result of these terms, along with the relative 
strengths of $\Delta_{ex}, \Delta_{v}$, and $m$, give a rich band dispersion 
and also account for the differences in the band dispersions of the two hetero-structures 
that we explore next.

The difference in the band dispersions of the two hetero-structures is 
related to the relative magnitudes of $\Delta_{ex}$, $\Delta_v$, and $m$. 
The energy dispersion of model Hamiltonian $H_{0}$ is
\begin{equation} \label{eq:Solution}
E_{\pm} = \pm \Delta_{ex} \pm \sqrt{m^2+v_{F}^2 |p|^2+\Delta_v^2 \pm 2 \Delta_v \sqrt{m^2+v^2 p_y^2}},
\end{equation}
where $|p| = \sqrt{p_{x}^2+p_{y}^2}$. 
For $\Delta_{ex} = 0$ the band dispersion has two important features, 
if $ m \geq \Delta_{v}$ the dispersion is elliptical and gapped at 
${\bf \Gamma}$ ($p_{x}=p_{y}=0$), 
with an energy gap $2|\Delta_{v} -m|$. 
In contrast when $\Delta_{v} > m$ the Dirac points shift from 
${\bf \Gamma}$ to $(0,\pm\sqrt{\Delta_{v}^2-m^2}/v_{F})$ 
and graphene retains its semi-metallic structure 
with two Dirac cones at $(0,\pm\sqrt{\Delta_{v}^2-m^2}/v_{F})$. 
%
%
%
%
%
%
%
For $\Delta_{ex} \neq 0$ and  $m  \geq \Delta_{v}$, 
there are three possibilities determined by the relative magnitudes 
of $\Delta_{ex}$ and $|\Delta_v-m|$. 
When (a) $\Delta_{ex} < |m-\Delta_v|$, 
there is a clear 
gap between the spin resolved states in Fig. \ref{fig:Configuration_MORE}(a). 
For (b) $\Delta_{ex} > |m-\Delta_v|$, 
the band dispersion exhibits an overlap between spin-up and spin-down bands 
shown in Fig. \ref{fig:Configuration_MORE}(b). 
Finally, at the transition point between scenarios (a) and (b) when 
(c) $\Delta_{ex} = |m-\Delta_v|$, 
the elliptical bands touch. 
This indicates that when $m  \geq \Delta_{v}$ 
there a is critical value of $\Delta_{ex} > |m- \Delta_{v}|$ 
at which the spin resolved states intersect.
On the other hand, if $\Delta_{ex} \neq 0$ and $\Delta_{v} > m$, the shifted Dirac 
points which now appear at $(0,\pm\sqrt{\Delta_{v}^2-m^2}/v_{F})$ exhibit 
crossing of spin-resolved bands 
indicating that spin-resolved bands cross for any value of $\Delta_{ex} \neq 0$. 
The gap at {$\bf \Gamma$} also depends on the relation 
between $\Delta_{ex}$ and $|\Delta_v-m|$. 
When $\Delta_{ex} < |\Delta_v-m|$, 
the band structure is shown in Fig. \ref{fig:Configuration_MORE}(c), 
and the case of $\Delta_{ex} > |\Delta_v-m|$ is shown in Fig. \ref{fig:Configuration_MORE}(d).
Now that we have established the conditions for the intersection of spin resolved bands, 
we explore the results of spin-orbit coupling on the graphene/EuO hetero-structures.

\subsection{Spin Orbit Coupling}

Spin-orbit coupling introduces
two additional terms consistent with the lattice symmetries, 
\begin{equation} \label{eq:H_SOC}
H_{\rm SOC} = \frac{\lambda_R}{2}(\hat{\sigma}_x \hat{s}_y \hat{\tau}_z-\hat{\sigma}_y \hat{s}_x)+\lambda_I \hat{\sigma}_z \hat{\tau}_z.
\end{equation}
The first term is the Rashba spin-orbit coupling
which breaks inversion symmetry in the plane perpendicular to the graphene sheet. 
Hence, $\lambda_{R} = 0 $ in the aligned structure. 
The second term is the intrinsic spin-orbit term that breaks time reversal symmetry. 
Therefore $\lambda_{I} \neq 0$ for both structures. 
Since in-plane inversion symmetry is preserved in both structures, we 
neglect the Dresselhaus spin-orbit coupling. 
Our calculations indicate that the strength of the spin-orbit coupling 
represented by $\lambda_{R}$ and $\lambda_{I}$ is always smaller than $m$, 
$\Delta_{V}$ and $\Delta_{ex}$, so we restrict our discussions to this case. 
For $\lambda_{R}, \lambda_{I} < m, \Delta_{V}, \Delta_{ex}$, 
the spin-orbit coupling introduces gaps between spin-resolved bands whenever they intersect 
(for example see Fig. \ref{fig:Configuration_MORE} (b),(c) and (d)).
With the addition of spin-orbit coupling the band dispersion becomes gapped, 
and the bands are a linear combination of spin-up and spin-down states. 

\begin{figure}
\includegraphics[width=.5\textwidth]{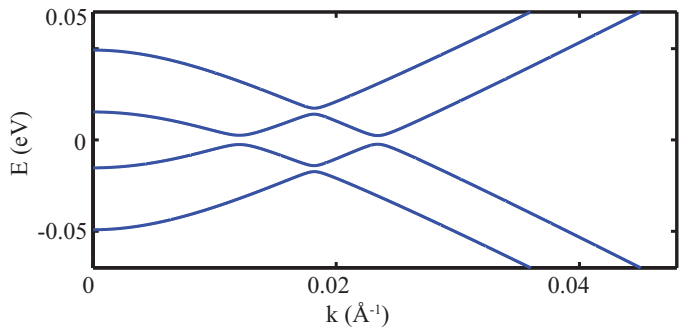}  
\caption{ 
Band dispersion with SOC of Eu-misaligned structure calculated from 
model Hamiltonian $H$ along the path ${\bf \Gamma}$ to ${\bf K}$ 
using parameters obtained from {\it ab initio} calculations. 
$\hbar v_F = 3.5$ eV$\cdot$ {\SI{}{\angstrom}}, 
$\Delta_{ex} = 80$ meV, $m = 48$ meV, $\Delta_v = 17$ meV, $\lambda_R = 5$ meV, 
and $\lambda_I = 1$ meV.
}
\label{fig:E-k_model_H_SOC} 
\end{figure}

By fitting the band dispersion with spin-orbit coupling 
obtained from the {\it ab initio} calculations shown in 
Fig. \ref{fig:SOC}, 
we determine the best fit parameters for our model Hamiltonian. 
This gives $\hbar v_F = 3.5$ eV$\cdot$ {\SI{}{\angstrom}}, 
$\Delta_{ex} = 80$ meV, $m = 48$ meV, $\Delta_v = 17$ meV, $\lambda_R = 5$ meV, 
and $\lambda_I = 1$ meV. 
The band dispersion 
along the path ${\bf \Gamma} - {\bf K}$ 
calculated from the model Hamiltonian $H = H_0 + H_{\rm SOC}$ is shown in 
Fig. \ref{fig:E-k_model_H_SOC}. 
The model Hamiltonian captures all four anti-crossing gaps at
about the same position in momentum space. 
Next, we study the topological properties of these bands obtained from $H$ and 
calculate the Hall conductance for a range of band parameters.

\section{Quantized anomalous Hall effect in graphene/EuO hetero-structures}

In Ref.~\onlinecite{qiao_quantum_2010}, Qiao {\it et. al.} 
found that ferromagnetic graphene in the presence of Rashba spin-orbit coupling 
shows the QAH effect with $ \sigma_{xy} = 2 e^2/h$. 
First principle calculations also demonstrated that this 
QAH phase can be engineered by doping with $3d$ or $5d$ transition-metal atoms 
or the proximity of a layered antiferromagentic insulator. 
In all cases studies thus far, 
the low energy bands are at the 
${\bf K}$ and ${\bf K}'$ points of the hexagonal BZ, 
and the Hall conductance in the gap is quantized $\sigma_{xy} = 2 e^2/h$ 
as long as $\lambda_{R} \neq 0$ and $\Delta_{ex} \neq 0$. 
In the graphene/EuO hetero-structure, as shown in sections II and III, 
the low-energy bands are no longer at ${\bf K}$ and ${\bf K}'$ but at ${\bf \Gamma}$,
and inter-valley interactions significantly modify the band dispersion 
and hence the topological properties of the bands. 
Therefore, we now analyze the effect of inter-valley interactions
on the topological
properties of graphene/EuO structures 
in the presence of in-plane inversion symmetry.

\begin{figure}
\centering
\includegraphics[width=.5\textwidth]{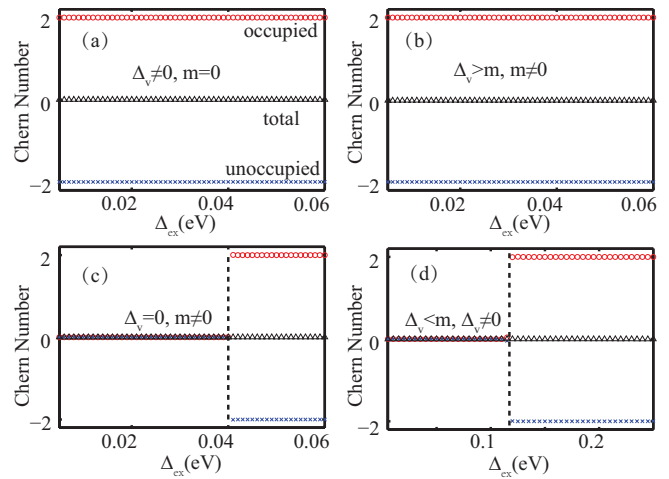}  
\caption{
(Color online)
The Chern number calculated as a function of $\Delta_{ex}$ for 4 different cases of 
$\Delta_v$ and $m$. 
The red open circles show the Chern number of the occupied bands, 
the blue `x' symbols show the Chern number of the unoccupied bands, 
and the black triangles show the Chern number of the summation of all bands.
The Chern number of the occupied bands is 2 for all values satisfying
(a) $\Delta_v \neq 0$, $m = 0$ and
(b) $\Delta_v > m$, $m \neq 0$. 
For condition (c), $\Delta_v = 0$,  $m \neq 0$,
the Chern number of the occupied bands becomes 2 
for $\Delta_{ex} \geq m$. For this example, $m$ is chosen to be 0.04 eV.
(d) For $\Delta_v \neq 0$ and $m > \Delta_v$, 
(in this example $\Delta_v = 0.01$ eV and $m = 0.04$ eV) 
the topological transition is pushed to a higher value of $\Delta_{ex} = 0.12$ eV.
}
\label{fig:Chern_4graph} 
\end{figure}

The Hall conductance is calculated from the  
integral of the Berry curvature over the BZ of the occupied bands and can be expressed as 
\begin{equation}
\label{sigmaint}
\sigma_{xy} = \frac{e^2}{\hbar} \sum_{\alpha} \int_{BZ} \frac{d^{2}p}{(2 \pi)^2} 
\Theta ( E_{F}- \epsilon_{\alpha}({\bf p}) ) \Omega_{\alpha} ({\bf p}),
\end{equation}
where $\alpha $ corresponds to the band index, $E_{F}$ denotes the Fermi energy, $\epsilon_{\alpha} ({\bf p})$
is the energy eigenstate, and $\Omega_{\alpha}({\bf p})$ is the Berry curvature of the 
$\alpha^{th}$ band. 
The Berry curvature in terms of the band eigenstates can be expressed as 
\begin{equation}
\label{BC_eq}
\Omega_{\alpha}({\bf p}) =  {\rm Im} \sum_{\beta \neq \alpha} \bigg[ \epsilon_{i j} 
\frac{\langle u_{\alpha} | \partial H ({\bf p})/\partial 
p_{i} | u_{\beta} \rangle \langle u_{\beta} |
\partial H ({\bf p})/\partial p_{j} | u_{\alpha}
\rangle}{(\epsilon_{\beta}({\bf p}) - \epsilon_{\alpha} ({\bf p}))^2}\bigg],
\end{equation}
where the Einstein summation convention is used for the roman indices $i$ and $j$,
$\epsilon_{ij}$ is anti-symmetric tensor, and 
$u_{\alpha} ({\bf p})$ is the $\alpha^{th}$ band eigenstate. 
It is instructive to note that 
in-plane inversion symmetry dictates $\Omega_{\alpha} ({\bf p}) = \Omega_{\alpha} (-{\bf p})$ 
and time reversal symmetry imposes  $\Omega_{\alpha}({\bf p}) = - \Omega_{\alpha} (-{\bf p})$. 
For graphene/EuO hetero-structures time reversal symmetry is broken due to exchange 
splitting caused by the ferromagnetic substrate, 
however in-plane inversion symmetry is preserved.
We take advantage of the in-plane inversion symmetry by calculating Berry curvature in the upper half-plane
$p_y > 0 $ and multiplying by a factor of $2$ to account for the lower half-plane $p_{y} <0$.
The Berry curvature is calculated numerically. 
Our calculations satisfy that 
the sum of the Berry curvatures over all the bands 
is zero at every ${\bf p}$ point in the BZ, as expected from Eq. (\ref{BC_eq}).

The model Hamiltonian $H$ only captures the low-energy bands near ${\bf \Gamma}$ and may not  
be valid over the full BZ of the graphene/EuO hetero-structure. 
The Berry curvature calculated using (\ref{BC_eq}) 
falls rapidly away from the ${\bf \Gamma}$ point. 
This allows us to  
restrict our calculations of the Hall conductance to a neighborhood of 
${\bf \Gamma}$. 
It is well known that when the Fermi energy lies in the gap $\sigma_{xy}$ is quantized 
and the Hall conductance at zero temperature can be expressed as
\begin{equation}
\sigma_{xy} = \frac{e^2}{h} \sum_{\alpha}^{\prime} C_{\alpha},
\end{equation}
where the prime indicates summation over the occupied bands, and
$C_{\alpha}$ is the Chern number of the $\alpha^{th}$ band 
that we calculate for different parameters of our model Hamiltonian $H $. 
We next discuss the Chern numbers at $E_F =0$ 
for the occupied and unoccupied bands.

The calculations for the Chern numbers were performed for 
4 different cases with fixed values for $\lambda_{R}$, $m$, and  $\Delta_v$ 
(with $\lambda_{R} < m, \; \Delta_v$) as a function of $\Delta_{ex}$ . 
The four different cases are depicted in Fig. \ref{fig:Chern_4graph}(a)-(d). 
Figs. \ref{fig:Chern_4graph}(a) and (b) show the results when $\Delta_v > m$. 
In this case the Chern number is quantized and gives 
a Hall conductance $\sigma_{xy} = 2 e^2/h$ for any value of $\lambda_R \neq 0$. 
However, when $\Delta_{v} < m$ and $m \neq 0$, 
there is a topological transition as a function of $\Delta_{ex}$ 
and the Chern number changes from $0$ to $2$ as shown in Figs.~\ref{fig:Chern_4graph}(c,d). 
For $\Delta_{v} = 0$, the transition occurs when $\Delta_{ex} \geq m$
as shown in Fig. \ref{fig:Chern_4graph}(c). 
For $\Delta_{v} \neq 0$ and $m > \Delta_v$, the transition is pushed to a higher value of
$\Delta_{ex}$ as shown in Fig. \ref{fig:Chern_4graph}(d).
For both (c) and (d), $m = 0.04$ eV. 
In (d), $\Delta_v = 0.01$ eV,
and the transition occurs at $\Delta_{ex} = 0.12$ eV. 
Unfortunately, we have been unable to find an analytical expression 
for the topological transition for $\Delta_{v} < m$. 
From the calculations, we conclude that the Chern number is 
$2$ for the case of $\Delta_v > m$ and that the 
system undergoes a topological transition for $\Delta_v < m$ as a function of $\Delta_{ex}$.

\begin{figure}
\centering
\includegraphics[width=.5\textwidth]{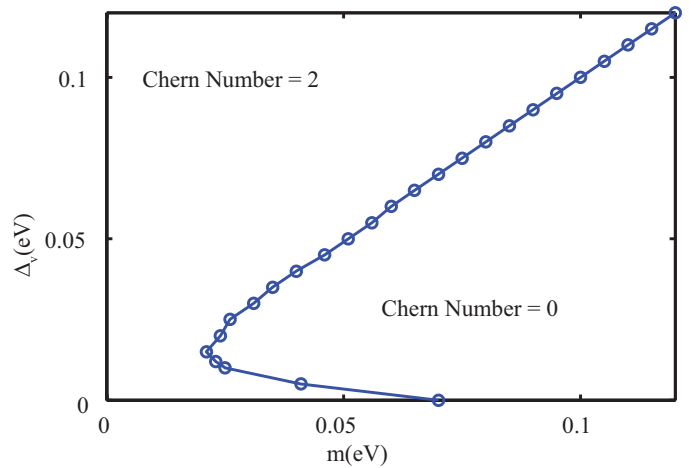}  
\caption{Phase diagram as a fucntion of $m$ and $\Delta_v$ for fixed 
values of $\lambda_{R} = 7$ meV and $\Delta_{ex}=70$ meV. 
The curve indicates the phase boundary between a
Chern number of
2 on the left side of the curve and
0 on the right side of the curve. 
}
\label{fig:phase} 
\end{figure}

The results can be summarized in terms of the phase diagram shown in Fig. \ref{fig:phase}. 
In this calculation, $\lambda_R = 7$ meV, and $\Delta_{ex} = 70$ meV are constant, 
and the 
behavior of Chern number is calculated as a function of both 
the magnitude of the valley pseudospin Zeeman term $\Delta_v$ and 
the inter-valley scattering term $m$.  
As shown in Fig. \ref{fig:phase}, 
when $\Delta_v > m$, the Chern number is always 2 giving a Hall conductance $\sigma_{xy} = 2 e^2/h$. 
At small values of $\Delta_v$,
a more complicated situation occurs in the region 
$m > \Delta_v$ of the phase diagram, however, 
for sufficiently large values of $\Delta_{v}$, the phase transition occurs at 
$m > \Delta_{v}$ for a fixed value of $\Delta_{ex}$.

\section{Conclusions and outlook}

Using insights from first principle calculations and lattice symmetries, 
we constructed a model Hamiltonian to describe commensurate graphene/EuO hetero-structures. 
In commensurate graphene/EuO structures band folding maps the Dirac cones to the 
$\bf{\Gamma}$ point of the hexagonal superlattice BZ of the combined hetero-structures.
Apart from inducing proximity exchange splitting to the graphene bands, the Eu atoms 
also introduce two distinct types of inter-valley interactions,
a valley pseudospin Zeeman term and an inter-valley scattering term, 
whose strengths are captured by two model parameters 
$\Delta_{v}$ and $m$. 
The combined effect of exchange and inter-valley interactions 
results in a non-linear dispersion at the $\bf{\Gamma}$
point which is captured by the model Hamiltonian. 
The parameters of the model Hamiltonian are determined by fitting to 
the band dispersion obtained from the {\it ab initio}
calculations.  

Using the model Hamiltonian with Rashba SOC, exchange, and inter-valley interactions we calculate the band 
dispersion and the topological properties of the commensurate graphene/EuO hetero-structures. 
The inter-valley interactions can significantly influence the topological properties of the bands for 
non-zero Rashba ($\lambda_{R} \neq 0$) and exchange splitting ($\Delta_{ex} \neq 0$).
For $\Delta_{v} > m$ with $\lambda_{R},\;\Delta_{ex} \neq 0$, 
the commensurate graphene/EuO heterostructure is a Chern insulator with a 
Hall conductance $\sigma_{xy} = 2e^2/h$, 
whereas for small $\Delta_v$ with $m> \Delta_{v}$, 
the phase diagram becomes more complicated and one needs 
a large exchange splitting or Rashba SOC to realize the Chern insulating phase.
Our calculations indicate that even in the presence of in-plane inversion symmetry, 
inter-valley interactions can significantly influence the topological 
properties of graphene/EuO hetero-structures. 

For a random incommensurate crystallographic stacking of graphene on EuO, 
the inter-valley coupling will be negligible since the Dirac cones will remain at $\bf K$ and $\bf K'$. 
However, since any incommensurate stacking will break the in-plane inversion symmetry of the graphene layer, 
the model Hamiltonian $H$ would acquire an additional term $M \hat{\sigma}_{z}$ \cite{ren_single-valley_2015}. 
In this case the topological properties will depend on the 
relative strength of $M$ and $\Delta_{gap} < \min(\lambda_{R},\Delta_{ex})$. 
The system will exhibit a QAH effect with a Chern number $2$ only if $M < \Delta_{gap}$. 

To observe the QAH effect in graphene/ferromagnet hetero-structures, 
it is important that the disorder induced broadening $\Sigma$ of 
the bands be smaller than the topological band gap $\Delta_{gap}$. 
The critical temperature required to observe the QAH effect is proportional to the mobility gap 
defined as $\Delta_{gap} - \Sigma $, which must be positive.
In order to increase the topological gap $\Delta_{gap}$, 
it is important to have a large Rashba SOC $\lambda_R$, 
which is small $\sim 7$ meV in our calculations. 
The Rashba spin-orbit coupling can in principle be enhanced by 
hydrogenation or deposition of heavy adatoms on the graphene surface\cite{balakrishnan_colossal_2013}. 
Alternatively, the spin orbit coupling can also be enhanced 
by creating hetero-structures consisting of transition metal 
dichalcogenides (TMDs) which possess a large spin-orbit coupling 
(e.g. TMD/graphene/EuO)\cite{avsar_spinorbit_2014}. 
Even for negative values of the mobility gap ($\Delta_{gap} - \Sigma  < 0$), 
the graphene/ferromagnetic structures will exhibit 
an unquantized anomalous Hall effect. 
However, in this case the anomalous hall effect will be additionally influenced by 
disorder induced extrinsic effects \cite{nagaosa_anomalous_2010} like side-jump and skew scattering mechanisms which are beyond 
the scope of this study.
                
\noindent
{\em Acknowledgements:} 
This work was supported as part of the Spins and Heat in Nanoscale Electronic Systems (SHINES) 
an Energy Frontier Research Center funded by the U.S. Department of Energy, 
Office of Science, Basic Energy Sciences under Award \#DE-SC0012670. 
{\it Ab initio} calculations were also supported by
FAME, one of six centers of STARnet, a Semiconductor
Research Corporation program sponsored by MARCO and DARPA and
used the Extreme Science and Engineering Discovery Environment (XSEDE), 
which is supported by the NSF grant OCI-1053575.
 

\begin{thebibliography}{27}%
\makeatletter
\providecommand \@ifxundefined [1]{%
 \@ifx{#1\undefined}
}%
\providecommand \@ifnum [1]{%
 \ifnum #1\expandafter \@firstoftwo
 \else \expandafter \@secondoftwo
 \fi
}%
\providecommand \@ifx [1]{%
 \ifx #1\expandafter \@firstoftwo
 \else \expandafter \@secondoftwo
 \fi
}%
\providecommand \natexlab [1]{#1}%
\providecommand \enquote  [1]{``#1''}%
\providecommand \bibnamefont  [1]{#1}%
\providecommand \bibfnamefont [1]{#1}%
\providecommand \citenamefont [1]{#1}%
\providecommand \href@noop [0]{\@secondoftwo}%
\providecommand \href [0]{\begingroup \@sanitize@url \@href}%
\providecommand \@href[1]{\@@startlink{#1}\@@href}%
\providecommand \@@href[1]{\endgroup#1\@@endlink}%
\providecommand \@sanitize@url [0]{\catcode `\\12\catcode `\$12\catcode
  `\&12\catcode `\#12\catcode `\^12\catcode `\_12\catcode `\%12\relax}%
\providecommand \@@startlink[1]{}%
\providecommand \@@endlink[0]{}%
\providecommand \url  [0]{\begingroup\@sanitize@url \@url }%
\providecommand \@url [1]{\endgroup\@href {#1}{\urlprefix }}%
\providecommand \urlprefix  [0]{URL }%
\providecommand \Eprint [0]{\href }%
\providecommand \doibase [0]{http://dx.doi.org/}%
\providecommand \selectlanguage [0]{\@gobble}%
\providecommand \bibinfo  [0]{\@secondoftwo}%
\providecommand \bibfield  [0]{\@secondoftwo}%
\providecommand \translation [1]{[#1]}%
\providecommand \BibitemOpen [0]{}%
\providecommand \bibitemStop [0]{}%
\providecommand \bibitemNoStop [0]{.\EOS\space}%
\providecommand \EOS [0]{\spacefactor3000\relax}%
\providecommand \BibitemShut  [1]{\csname bibitem#1\endcsname}%
\let\auto@bib@innerbib\@empty
\bibitem [{\citenamefont {Thouless}\ \emph {et~al.}(1982)\citenamefont
  {Thouless}, \citenamefont {Kohmoto}, \citenamefont {Nightingale},\ and\
  \citenamefont {den Nijs}}]{TKNN}%
  \BibitemOpen
  \bibfield  {author} {\bibinfo {author} {\bibfnamefont {D.~J.}\ \bibnamefont
  {Thouless}}, \bibinfo {author} {\bibfnamefont {M.}~\bibnamefont {Kohmoto}},
  \bibinfo {author} {\bibfnamefont {M.~P.}\ \bibnamefont {Nightingale}}, \ and\
  \bibinfo {author} {\bibfnamefont {M.}~\bibnamefont {den Nijs}},\ }\href
  {\doibase 10.1103/PhysRevLett.49.405} {\bibfield  {journal} {\bibinfo
  {journal} {Phys. Rev. Lett.}\ }\textbf {\bibinfo {volume} {49}},\ \bibinfo
  {pages} {405} (\bibinfo {year} {1982})}\BibitemShut {NoStop}%
\bibitem [{\citenamefont {Haldane}(1988)}]{HaldaneQAH}%
  \BibitemOpen
  \bibfield  {author} {\bibinfo {author} {\bibfnamefont {F.~D.~M.}\
  \bibnamefont {Haldane}},\ }\href {\doibase 10.1103/PhysRevLett.61.2015}
  {\bibfield  {journal} {\bibinfo  {journal} {Phys. Rev. Lett.}\ }\textbf
  {\bibinfo {volume} {61}},\ \bibinfo {pages} {2015} (\bibinfo {year}
  {1988})}\BibitemShut {NoStop}%
\bibitem [{\citenamefont {Liu}\ \emph {et~al.}(2008{\natexlab{a}})\citenamefont
  {Liu}, \citenamefont {Qi}, \citenamefont {Dai}, \citenamefont {Fang},\ and\
  \citenamefont {Zhang}}]{HgQAH}%
  \BibitemOpen
  \bibfield  {author} {\bibinfo {author} {\bibfnamefont {C.-X.}\ \bibnamefont
  {Liu}}, \bibinfo {author} {\bibfnamefont {X.-L.}\ \bibnamefont {Qi}},
  \bibinfo {author} {\bibfnamefont {X.}~\bibnamefont {Dai}}, \bibinfo {author}
  {\bibfnamefont {Z.}~\bibnamefont {Fang}}, \ and\ \bibinfo {author}
  {\bibfnamefont {S.-C.}\ \bibnamefont {Zhang}},\ }\href {\doibase
  10.1103/PhysRevLett.101.146802} {\bibfield  {journal} {\bibinfo  {journal}
  {Phys. Rev. Lett.}\ }\textbf {\bibinfo {volume} {101}},\ \bibinfo {pages}
  {146802} (\bibinfo {year} {2008}{\natexlab{a}})}\BibitemShut {NoStop}%
\bibitem [{\citenamefont {Wu}(2008)}]{CWQAH}%
  \BibitemOpen
  \bibfield  {author} {\bibinfo {author} {\bibfnamefont {C.}~\bibnamefont
  {Wu}},\ }\href {\doibase 10.1103/PhysRevLett.101.186807} {\bibfield
  {journal} {\bibinfo  {journal} {Phys. Rev. Lett.}\ }\textbf {\bibinfo
  {volume} {101}},\ \bibinfo {pages} {186807} (\bibinfo {year}
  {2008})}\BibitemShut {NoStop}%
\bibitem [{\citenamefont {Liu}\ \emph {et~al.}(2008{\natexlab{b}})\citenamefont
  {Liu}, \citenamefont {Qi}, \citenamefont {Dai}, \citenamefont {Fang},\ and\
  \citenamefont {Zhang}}]{disorderedAIQAH}%
  \BibitemOpen
  \bibfield  {author} {\bibinfo {author} {\bibfnamefont {C.-X.}\ \bibnamefont
  {Liu}}, \bibinfo {author} {\bibfnamefont {X.-L.}\ \bibnamefont {Qi}},
  \bibinfo {author} {\bibfnamefont {X.}~\bibnamefont {Dai}}, \bibinfo {author}
  {\bibfnamefont {Z.}~\bibnamefont {Fang}}, \ and\ \bibinfo {author}
  {\bibfnamefont {S.-C.}\ \bibnamefont {Zhang}},\ }\href {\doibase
  10.1103/PhysRevLett.101.146802} {\bibfield  {journal} {\bibinfo  {journal}
  {Phys. Rev. Lett.}\ }\textbf {\bibinfo {volume} {101}},\ \bibinfo {pages}
  {146802} (\bibinfo {year} {2008}{\natexlab{b}})}\BibitemShut {NoStop}%
\bibitem [{\citenamefont {Yu}\ \emph {et~al.}(2010)\citenamefont {Yu},
  \citenamefont {Zhang}, \citenamefont {Zhang}, \citenamefont {Zhang},
  \citenamefont {Dai},\ and\ \citenamefont {Fang}}]{yu_quantized_2010}%
  \BibitemOpen
  \bibfield  {author} {\bibinfo {author} {\bibfnamefont {R.}~\bibnamefont
  {Yu}}, \bibinfo {author} {\bibfnamefont {W.}~\bibnamefont {Zhang}}, \bibinfo
  {author} {\bibfnamefont {H.-J.}\ \bibnamefont {Zhang}}, \bibinfo {author}
  {\bibfnamefont {S.-C.}\ \bibnamefont {Zhang}}, \bibinfo {author}
  {\bibfnamefont {X.}~\bibnamefont {Dai}}, \ and\ \bibinfo {author}
  {\bibfnamefont {Z.}~\bibnamefont {Fang}},\ }\href {\doibase
  10.1126/science.1187485} {\bibfield  {journal} {\bibinfo  {journal}
  {Science}\ }\textbf {\bibinfo {volume} {329}},\ \bibinfo {pages} {61}
  (\bibinfo {year} {2010})}\BibitemShut {NoStop}%
\bibitem [{\citenamefont {Jiang}\ \emph {et~al.}(2012)\citenamefont {Jiang},
  \citenamefont {Qiao}, \citenamefont {Liu},\ and\ \citenamefont
  {Niu}}]{jiang_quantum_2012}%
  \BibitemOpen
  \bibfield  {author} {\bibinfo {author} {\bibfnamefont {H.}~\bibnamefont
  {Jiang}}, \bibinfo {author} {\bibfnamefont {Z.}~\bibnamefont {Qiao}},
  \bibinfo {author} {\bibfnamefont {H.}~\bibnamefont {Liu}}, \ and\ \bibinfo
  {author} {\bibfnamefont {Q.}~\bibnamefont {Niu}},\ }\href {\doibase
  10.1103/PhysRevB.85.045445} {\bibfield  {journal} {\bibinfo  {journal} {Phys.
  Rev. B}\ }\textbf {\bibinfo {volume} {85}},\ \bibinfo {pages} {045445}
  (\bibinfo {year} {2012})}\BibitemShut {NoStop}%
\bibitem [{\citenamefont {Qiao}\ \emph {et~al.}(2010)\citenamefont {Qiao},
  \citenamefont {Yang}, \citenamefont {Feng}, \citenamefont {Tse},
  \citenamefont {Ding}, \citenamefont {Yao}, \citenamefont {Wang},\ and\
  \citenamefont {Niu}}]{qiao_quantum_2010}%
  \BibitemOpen
  \bibfield  {author} {\bibinfo {author} {\bibfnamefont {Z.}~\bibnamefont
  {Qiao}}, \bibinfo {author} {\bibfnamefont {S.~A.}\ \bibnamefont {Yang}},
  \bibinfo {author} {\bibfnamefont {W.}~\bibnamefont {Feng}}, \bibinfo {author}
  {\bibfnamefont {W.-K.}\ \bibnamefont {Tse}}, \bibinfo {author} {\bibfnamefont
  {J.}~\bibnamefont {Ding}}, \bibinfo {author} {\bibfnamefont {Y.}~\bibnamefont
  {Yao}}, \bibinfo {author} {\bibfnamefont {J.}~\bibnamefont {Wang}}, \ and\
  \bibinfo {author} {\bibfnamefont {Q.}~\bibnamefont {Niu}},\ }\href {\doibase
  10.1103/PhysRevB.82.161414} {\bibfield  {journal} {\bibinfo  {journal} {Phys.
  Rev. B}\ }\textbf {\bibinfo {volume} {82}},\ \bibinfo {pages} {161414}
  (\bibinfo {year} {2010})}\BibitemShut {NoStop}%
\bibitem [{\citenamefont {Qiao}\ \emph {et~al.}(2012)\citenamefont {Qiao},
  \citenamefont {Jiang}, \citenamefont {Li}, \citenamefont {Yao},\ and\
  \citenamefont {Niu}}]{Niu_QAH_longpaper}%
  \BibitemOpen
  \bibfield  {author} {\bibinfo {author} {\bibfnamefont {Z.}~\bibnamefont
  {Qiao}}, \bibinfo {author} {\bibfnamefont {H.}~\bibnamefont {Jiang}},
  \bibinfo {author} {\bibfnamefont {X.}~\bibnamefont {Li}}, \bibinfo {author}
  {\bibfnamefont {Y.}~\bibnamefont {Yao}}, \ and\ \bibinfo {author}
  {\bibfnamefont {Q.}~\bibnamefont {Niu}},\ }\href {\doibase
  10.1103/PhysRevB.85.115439} {\bibfield  {journal} {\bibinfo  {journal} {Phys.
  Rev. B}\ }\textbf {\bibinfo {volume} {85}},\ \bibinfo {pages} {115439}
  (\bibinfo {year} {2012})}\BibitemShut {NoStop}%
\bibitem [{\citenamefont {Hasan}\ and\ \citenamefont {Kane}(2010)}]{TIreview}%
  \BibitemOpen
  \bibfield  {author} {\bibinfo {author} {\bibfnamefont {M.~Z.}\ \bibnamefont
  {Hasan}}\ and\ \bibinfo {author} {\bibfnamefont {C.~L.}\ \bibnamefont
  {Kane}},\ }\href {\doibase 10.1103/RevModPhys.82.3045} {\bibfield  {journal}
  {\bibinfo  {journal} {Rev. Mod. Phys.}\ }\textbf {\bibinfo {volume} {82}},\
  \bibinfo {pages} {3045} (\bibinfo {year} {2010})}\BibitemShut {NoStop}%
\bibitem [{\citenamefont {Kou}\ \emph {et~al.}(2014)\citenamefont {Kou},
  \citenamefont {Guo}, \citenamefont {Fan}, \citenamefont {Pan}, \citenamefont
  {Lang}, \citenamefont {Jiang}, \citenamefont {Shao}, \citenamefont {Nie},
  \citenamefont {Murata}, \citenamefont {Tang}, \citenamefont {Wang},
  \citenamefont {He}, \citenamefont {Lee}, \citenamefont {Lee},\ and\
  \citenamefont {Wang}}]{kou_scale-invariant_2014}%
  \BibitemOpen
  \bibfield  {author} {\bibinfo {author} {\bibfnamefont {X.}~\bibnamefont
  {Kou}}, \bibinfo {author} {\bibfnamefont {S.-T.}\ \bibnamefont {Guo}},
  \bibinfo {author} {\bibfnamefont {Y.}~\bibnamefont {Fan}}, \bibinfo {author}
  {\bibfnamefont {L.}~\bibnamefont {Pan}}, \bibinfo {author} {\bibfnamefont
  {M.}~\bibnamefont {Lang}}, \bibinfo {author} {\bibfnamefont {Y.}~\bibnamefont
  {Jiang}}, \bibinfo {author} {\bibfnamefont {Q.}~\bibnamefont {Shao}},
  \bibinfo {author} {\bibfnamefont {T.}~\bibnamefont {Nie}}, \bibinfo {author}
  {\bibfnamefont {K.}~\bibnamefont {Murata}}, \bibinfo {author} {\bibfnamefont
  {J.}~\bibnamefont {Tang}}, \bibinfo {author} {\bibfnamefont {Y.}~\bibnamefont
  {Wang}}, \bibinfo {author} {\bibfnamefont {L.}~\bibnamefont {He}}, \bibinfo
  {author} {\bibfnamefont {T.-K.}\ \bibnamefont {Lee}}, \bibinfo {author}
  {\bibfnamefont {W.-L.}\ \bibnamefont {Lee}}, \ and\ \bibinfo {author}
  {\bibfnamefont {K.~L.}\ \bibnamefont {Wang}},\ }\href {\doibase
  10.1103/PhysRevLett.113.137201} {\bibfield  {journal} {\bibinfo  {journal}
  {Phys. Rev. Lett.}\ }\textbf {\bibinfo {volume} {113}},\ \bibinfo {pages}
  {137201} (\bibinfo {year} {2014})}\BibitemShut {NoStop}%
\bibitem [{\citenamefont {Wang}\ \emph {et~al.}(2015)\citenamefont {Wang},
  \citenamefont {Tang}, \citenamefont {Sachs}, \citenamefont {Barlas},\ and\
  \citenamefont {Shi}}]{ShiPRL}%
  \BibitemOpen
  \bibfield  {author} {\bibinfo {author} {\bibfnamefont {Z.}~\bibnamefont
  {Wang}}, \bibinfo {author} {\bibfnamefont {C.}~\bibnamefont {Tang}}, \bibinfo
  {author} {\bibfnamefont {R.}~\bibnamefont {Sachs}}, \bibinfo {author}
  {\bibfnamefont {Y.}~\bibnamefont {Barlas}}, \ and\ \bibinfo {author}
  {\bibfnamefont {J.}~\bibnamefont {Shi}},\ }\href {\doibase
  10.1103/PhysRevLett.114.016603} {\bibfield  {journal} {\bibinfo  {journal}
  {Phys. Rev. Lett.}\ }\textbf {\bibinfo {volume} {114}},\ \bibinfo {pages}
  {016603} (\bibinfo {year} {2015})}\BibitemShut {NoStop}%
\bibitem [{\citenamefont {Qiao}\ \emph {et~al.}(2014)\citenamefont {Qiao},
  \citenamefont {Ren}, \citenamefont {Chen}, \citenamefont {Bellaiche},
  \citenamefont {Zhang}, \citenamefont {MacDonald},\ and\ \citenamefont
  {Niu}}]{qiao_quantum_2014}%
  \BibitemOpen
  \bibfield  {author} {\bibinfo {author} {\bibfnamefont {Z.}~\bibnamefont
  {Qiao}}, \bibinfo {author} {\bibfnamefont {W.}~\bibnamefont {Ren}}, \bibinfo
  {author} {\bibfnamefont {H.}~\bibnamefont {Chen}}, \bibinfo {author}
  {\bibfnamefont {L.}~\bibnamefont {Bellaiche}}, \bibinfo {author}
  {\bibfnamefont {Z.}~\bibnamefont {Zhang}}, \bibinfo {author} {\bibfnamefont
  {A.~H.}\ \bibnamefont {MacDonald}}, \ and\ \bibinfo {author} {\bibfnamefont
  {Q.}~\bibnamefont {Niu}},\ }\href {\doibase 10.1103/PhysRevLett.112.116404}
  {\bibfield  {journal} {\bibinfo  {journal} {Phys. Rev. Lett.}\ }\textbf
  {\bibinfo {volume} {112}},\ \bibinfo {pages} {116404} (\bibinfo {year}
  {2014})}\BibitemShut {NoStop}%
\bibitem [{\citenamefont {Yang}\ \emph {et~al.}(2013)\citenamefont {Yang},
  \citenamefont {Hallal}, \citenamefont {Terrade}, \citenamefont {Waintal},
  \citenamefont {Roche},\ and\ \citenamefont {Chshiev}}]{yang_proximity_2013}%
  \BibitemOpen
  \bibfield  {author} {\bibinfo {author} {\bibfnamefont {H.~X.}\ \bibnamefont
  {Yang}}, \bibinfo {author} {\bibfnamefont {A.}~\bibnamefont {Hallal}},
  \bibinfo {author} {\bibfnamefont {D.}~\bibnamefont {Terrade}}, \bibinfo
  {author} {\bibfnamefont {X.}~\bibnamefont {Waintal}}, \bibinfo {author}
  {\bibfnamefont {S.}~\bibnamefont {Roche}}, \ and\ \bibinfo {author}
  {\bibfnamefont {M.}~\bibnamefont {Chshiev}},\ }\href {\doibase
  10.1103/PhysRevLett.110.046603} {\bibfield  {journal} {\bibinfo  {journal}
  {Phys. Rev. Lett.}\ }\textbf {\bibinfo {volume} {110}},\ \bibinfo {pages}
  {046603} (\bibinfo {year} {2013})}\BibitemShut {NoStop}%
\bibitem [{\citenamefont {Qi}\ \emph {et~al.}(2015)\citenamefont {Qi},
  \citenamefont {Li}, \citenamefont {Niu},\ and\ \citenamefont
  {Feng}}]{qi_giant_2015}%
  \BibitemOpen
  \bibfield  {author} {\bibinfo {author} {\bibfnamefont {J.}~\bibnamefont
  {Qi}}, \bibinfo {author} {\bibfnamefont {X.}~\bibnamefont {Li}}, \bibinfo
  {author} {\bibfnamefont {Q.}~\bibnamefont {Niu}}, \ and\ \bibinfo {author}
  {\bibfnamefont {J.}~\bibnamefont {Feng}},\ }\href
  {http://arxiv.org/abs/1504.04434} {\  (\bibinfo {year} {2015})},\ \Eprint
  {http://arxiv.org/abs/1504.04434} {1504.04434} \BibitemShut {NoStop}%
\bibitem [{\citenamefont {Kresse}\ and\ \citenamefont
  {Furthm{\"u}ller}(1996{\natexlab{a}})}]{kresse_efficient_1996}%
  \BibitemOpen
  \bibfield  {author} {\bibinfo {author} {\bibfnamefont {G.}~\bibnamefont
  {Kresse}}\ and\ \bibinfo {author} {\bibfnamefont {J.}~\bibnamefont
  {Furthm{\"u}ller}},\ }\href {\doibase 10.1103/PhysRevB.54.11169} {\bibfield
  {journal} {\bibinfo  {journal} {Phys. Rev. B}\ }\textbf {\bibinfo {volume}
  {54}},\ \bibinfo {pages} {11169} (\bibinfo {year}
  {1996}{\natexlab{a}})}\BibitemShut {NoStop}%
\bibitem [{\citenamefont {Kresse}\ and\ \citenamefont
  {Hafner}(1993)}]{kresse_ab_1993}%
  \BibitemOpen
  \bibfield  {author} {\bibinfo {author} {\bibfnamefont {G.}~\bibnamefont
  {Kresse}}\ and\ \bibinfo {author} {\bibfnamefont {J.}~\bibnamefont
  {Hafner}},\ }\href {\doibase 10.1103/PhysRevB.47.558} {\bibfield  {journal}
  {\bibinfo  {journal} {Phys. Rev. B}\ }\textbf {\bibinfo {volume} {47}},\
  \bibinfo {pages} {558} (\bibinfo {year} {1993})}\BibitemShut {NoStop}%
\bibitem [{\citenamefont {Kresse}\ and\ \citenamefont
  {Furthm{\"u}ller}(1996{\natexlab{b}})}]{kresse_efficiency_1996}%
  \BibitemOpen
  \bibfield  {author} {\bibinfo {author} {\bibfnamefont {G.}~\bibnamefont
  {Kresse}}\ and\ \bibinfo {author} {\bibfnamefont {J.}~\bibnamefont
  {Furthm{\"u}ller}},\ }\href {\doibase 10.1016/0927-0256(96)00008-0}
  {\bibfield  {journal} {\bibinfo  {journal} {Computational Materials Science}\
  }\textbf {\bibinfo {volume} {6}},\ \bibinfo {pages} {15} (\bibinfo {year}
  {1996}{\natexlab{b}})}\BibitemShut {NoStop}%
\bibitem [{\citenamefont {Bl{\"o}chl}(1994)}]{blochl_projector_1994}%
  \BibitemOpen
  \bibfield  {author} {\bibinfo {author} {\bibfnamefont {P.~E.}\ \bibnamefont
  {Bl{\"o}chl}},\ }\href {\doibase 10.1103/PhysRevB.50.17953} {\bibfield
  {journal} {\bibinfo  {journal} {Phys. Rev. B}\ }\textbf {\bibinfo {volume}
  {50}},\ \bibinfo {pages} {17953} (\bibinfo {year} {1994})}\BibitemShut
  {NoStop}%
\bibitem [{\citenamefont {Perdew}\ \emph {et~al.}(1992)\citenamefont {Perdew},
  \citenamefont {Chevary}, \citenamefont {Vosko}, \citenamefont {Jackson},
  \citenamefont {Pederson}, \citenamefont {Singh},\ and\ \citenamefont
  {Fiolhais}}]{perdew_atoms_1992}%
  \BibitemOpen
  \bibfield  {author} {\bibinfo {author} {\bibfnamefont {J.~P.}\ \bibnamefont
  {Perdew}}, \bibinfo {author} {\bibfnamefont {J.~A.}\ \bibnamefont {Chevary}},
  \bibinfo {author} {\bibfnamefont {S.~H.}\ \bibnamefont {Vosko}}, \bibinfo
  {author} {\bibfnamefont {K.~A.}\ \bibnamefont {Jackson}}, \bibinfo {author}
  {\bibfnamefont {M.~R.}\ \bibnamefont {Pederson}}, \bibinfo {author}
  {\bibfnamefont {D.~J.}\ \bibnamefont {Singh}}, \ and\ \bibinfo {author}
  {\bibfnamefont {C.}~\bibnamefont {Fiolhais}},\ }\href {\doibase
  10.1103/PhysRevB.46.6671} {\bibfield  {journal} {\bibinfo  {journal} {Phys.
  Rev. B}\ }\textbf {\bibinfo {volume} {46}},\ \bibinfo {pages} {6671}
  (\bibinfo {year} {1992})}\BibitemShut {NoStop}%
\bibitem [{\citenamefont {Wang}\ and\ \citenamefont
  {Perdew}(1991)}]{wang_correlation_1991}%
  \BibitemOpen
  \bibfield  {author} {\bibinfo {author} {\bibfnamefont {Y.}~\bibnamefont
  {Wang}}\ and\ \bibinfo {author} {\bibfnamefont {J.~P.}\ \bibnamefont
  {Perdew}},\ }\href {\doibase 10.1103/PhysRevB.44.13298} {\bibfield  {journal}
  {\bibinfo  {journal} {Phys. Rev. B}\ }\textbf {\bibinfo {volume} {44}},\
  \bibinfo {pages} {13298} (\bibinfo {year} {1991})}\BibitemShut {NoStop}%
\bibitem [{\citenamefont {Kresse}\ and\ \citenamefont
  {Joubert}(1999)}]{kresse_ultrasoft_1999}%
  \BibitemOpen
  \bibfield  {author} {\bibinfo {author} {\bibfnamefont {G.}~\bibnamefont
  {Kresse}}\ and\ \bibinfo {author} {\bibfnamefont {D.}~\bibnamefont
  {Joubert}},\ }\href {\doibase 10.1103/PhysRevB.59.1758} {\bibfield  {journal}
  {\bibinfo  {journal} {Phys. Rev. B}\ }\textbf {\bibinfo {volume} {59}},\
  \bibinfo {pages} {1758} (\bibinfo {year} {1999})}\BibitemShut {NoStop}%
\bibitem [{\citenamefont {Ingle}\ and\ \citenamefont
  {Elfimov}(2008)}]{ingle_influence_2008}%
  \BibitemOpen
  \bibfield  {author} {\bibinfo {author} {\bibfnamefont {N.~J.~C.}\
  \bibnamefont {Ingle}}\ and\ \bibinfo {author} {\bibfnamefont {I.~S.}\
  \bibnamefont {Elfimov}},\ }\href {\doibase 10.1103/PhysRevB.77.121202}
  {\bibfield  {journal} {\bibinfo  {journal} {Phys. Rev. B}\ }\textbf {\bibinfo
  {volume} {77}},\ \bibinfo {pages} {121202} (\bibinfo {year}
  {2008})}\BibitemShut {NoStop}%
\bibitem [{\citenamefont {Ren}\ \emph {et~al.}(2015)\citenamefont {Ren},
  \citenamefont {Deng}, \citenamefont {Qiao}, \citenamefont {Li}, \citenamefont
  {Jung}, \citenamefont {Zeng}, \citenamefont {Zhang},\ and\ \citenamefont
  {Niu}}]{ren_single-valley_2015}%
  \BibitemOpen
  \bibfield  {author} {\bibinfo {author} {\bibfnamefont {Y.}~\bibnamefont
  {Ren}}, \bibinfo {author} {\bibfnamefont {X.}~\bibnamefont {Deng}}, \bibinfo
  {author} {\bibfnamefont {Z.}~\bibnamefont {Qiao}}, \bibinfo {author}
  {\bibfnamefont {C.}~\bibnamefont {Li}}, \bibinfo {author} {\bibfnamefont
  {J.}~\bibnamefont {Jung}}, \bibinfo {author} {\bibfnamefont {C.}~\bibnamefont
  {Zeng}}, \bibinfo {author} {\bibfnamefont {Z.}~\bibnamefont {Zhang}}, \ and\
  \bibinfo {author} {\bibfnamefont {Q.}~\bibnamefont {Niu}},\ }\href {\doibase
  10.1103/PhysRevB.91.245415} {\bibfield  {journal} {\bibinfo  {journal} {Phys.
  Rev. B}\ }\textbf {\bibinfo {volume} {91}},\ \bibinfo {pages} {245415}
  (\bibinfo {year} {2015})}\BibitemShut {NoStop}%
\bibitem [{\citenamefont {Balakrishnan}\ \emph {et~al.}(2013)\citenamefont
  {Balakrishnan}, \citenamefont {Kok Wai~Koon}, \citenamefont {Jaiswal},
  \citenamefont {Castro~Neto},\ and\ \citenamefont
  {{\"O}zyilmaz}}]{balakrishnan_colossal_2013}%
  \BibitemOpen
  \bibfield  {author} {\bibinfo {author} {\bibfnamefont {J.}~\bibnamefont
  {Balakrishnan}}, \bibinfo {author} {\bibfnamefont {G.}~\bibnamefont {Kok
  Wai~Koon}}, \bibinfo {author} {\bibfnamefont {M.}~\bibnamefont {Jaiswal}},
  \bibinfo {author} {\bibfnamefont {A.~H.}\ \bibnamefont {Castro~Neto}}, \ and\
  \bibinfo {author} {\bibfnamefont {B.}~\bibnamefont {{\"O}zyilmaz}},\ }\href
  {\doibase 10.1038/nphys2576} {\bibfield  {journal} {\bibinfo  {journal} {Nat
  Phys}\ }\textbf {\bibinfo {volume} {9}},\ \bibinfo {pages} {284} (\bibinfo
  {year} {2013})}\BibitemShut {NoStop}%
\bibitem [{\citenamefont {Avsar}\ \emph {et~al.}(2014)\citenamefont {Avsar},
  \citenamefont {Tan}, \citenamefont {Taychatanapat}, \citenamefont
  {Balakrishnan}, \citenamefont {Koon}, \citenamefont {Yeo}, \citenamefont
  {Lahiri}, \citenamefont {Carvalho}, \citenamefont {Rodin}, \citenamefont
  {O’Farrell}, \citenamefont {Eda}, \citenamefont {Castro~Neto},\ and\
  \citenamefont {{\"O}zyilmaz}}]{avsar_spinorbit_2014}%
  \BibitemOpen
  \bibfield  {author} {\bibinfo {author} {\bibfnamefont {A.}~\bibnamefont
  {Avsar}}, \bibinfo {author} {\bibfnamefont {J.~Y.}\ \bibnamefont {Tan}},
  \bibinfo {author} {\bibfnamefont {T.}~\bibnamefont {Taychatanapat}}, \bibinfo
  {author} {\bibfnamefont {J.}~\bibnamefont {Balakrishnan}}, \bibinfo {author}
  {\bibfnamefont {G.~K.~W.}\ \bibnamefont {Koon}}, \bibinfo {author}
  {\bibfnamefont {Y.}~\bibnamefont {Yeo}}, \bibinfo {author} {\bibfnamefont
  {J.}~\bibnamefont {Lahiri}}, \bibinfo {author} {\bibfnamefont
  {A.}~\bibnamefont {Carvalho}}, \bibinfo {author} {\bibfnamefont {A.~S.}\
  \bibnamefont {Rodin}}, \bibinfo {author} {\bibfnamefont {E.~C.~T.}\
  \bibnamefont {O’Farrell}}, \bibinfo {author} {\bibfnamefont
  {G.}~\bibnamefont {Eda}}, \bibinfo {author} {\bibfnamefont {A.~H.}\
  \bibnamefont {Castro~Neto}}, \ and\ \bibinfo {author} {\bibfnamefont
  {B.}~\bibnamefont {{\"O}zyilmaz}},\ }\href {\doibase 10.1038/ncomms5875}
  {\bibfield  {journal} {\bibinfo  {journal} {Nat Commun}\ }\textbf {\bibinfo
  {volume} {5}},\ \bibinfo {pages} {4875} (\bibinfo {year} {2014})}\BibitemShut
  {NoStop}%
\bibitem [{\citenamefont {Nagaosa}\ \emph {et~al.}(2010)\citenamefont
  {Nagaosa}, \citenamefont {Sinova}, \citenamefont {Onoda}, \citenamefont
  {{MacDonald}},\ and\ \citenamefont {Ong}}]{nagaosa_anomalous_2010}%
  \BibitemOpen
  \bibfield  {author} {\bibinfo {author} {\bibfnamefont {N.}~\bibnamefont
  {Nagaosa}}, \bibinfo {author} {\bibfnamefont {J.}~\bibnamefont {Sinova}},
  \bibinfo {author} {\bibfnamefont {S.}~\bibnamefont {Onoda}}, \bibinfo
  {author} {\bibfnamefont {A.~H.}\ \bibnamefont {{MacDonald}}}, \ and\ \bibinfo
  {author} {\bibfnamefont {N.~P.}\ \bibnamefont {Ong}},\ }\href {\doibase
  10.1103/RevModPhys.82.1539} {\bibfield  {journal} {\bibinfo  {journal} {Rev.
  Mod. Phys.}\ }\textbf {\bibinfo {volume} {82}},\ \bibinfo {pages} {1539}
  (\bibinfo {year} {2010})}\BibitemShut {NoStop}%
\end{thebibliography}
%

\end{document}